\documentclass[twocolumn,english]{revtex4-1}
\usepackage[T1]{fontenc}
\usepackage[latin9]{inputenc}
\usepackage{color}
\usepackage{babel}
\usepackage{units}
\usepackage{textcomp}
\usepackage{amsmath}
\usepackage{graphicx}
\usepackage{setspace}
\usepackage[unicode=true,pdfusetitle,
 bookmarks=true,bookmarksnumbered=false,bookmarksopen=false,
 breaklinks=false,pdfborder={0 0 0},pdfborderstyle={},backref=false,colorlinks=true]
 {hyperref}

\makeatletter

\providecommand{\tabularnewline}{\\}
\newcommand{\lyxdot}{.}

\usepackage{mathptmx}
\usepackage[caption=false]{subfig}
\newcommand*{\balancecolsandclearpage}{%
  \close@column@grid
  \cleardoublepage
  \twocolumngrid
}

\@ifundefined{showcaptionsetup}{}{%
 \PassOptionsToPackage{caption=false}{subfig}}
\usepackage{subfig}
\makeatother

\begin{document}

\title{Role of Spin, Phonon and Plasmon Dynamics on Ferromagnetism in Cr
doped 3C-SiC}

\author{Gyanti Prakash Moharana, Rahul Kothari}

\affiliation{Indian Institute of Technology Madras, Chennai-600036}

\author{S.K. Singh}

\affiliation{CSIR Innovation Center for Plasma Processing, IMMT Bhubaneshwar -
751013}

\author{N. Harish Kumar}
\email{nhk@iitm.ac.in}

\selectlanguage{english}%

\affiliation{Indian Institute of Technology Madras, Chennai - 600036}
\begin{abstract}
The defect induced magnetism has initiated a lot of interest in field
of spintronics. In this regard, SiC is a promising material because
of its unique properties under extreme conditions. Hence it will be
very much interesting to investigate the interaction between defects
and itinerant carrier in doped SiC for spintronics application. We
report the structural stability and magnetic interaction in Cr doped
3C-SiC synthesized by Thermal Plasma Technique. The EPR spectrum of
undoped 3C-SiC shows a sharp resonance line corresponding to $g=2.00$
associated with the defects present in the system. Anomalous temperature
evolution tendency of the relative intensity of EPR spectra can be
attributed to magnetically correlated defects in the host matrix.
For the first time we report the detailed quantitative analysis of
X-band and Q-band EPR study in Cr doped 3C-SiC which reveals that
Cr can be in multivalent state. The non monotonous variation of Longitudinal
Mode (LO) of the Raman spectra has been explained based on the interaction
between carriers and surface plasmon using Longitudinal optical plasmon
coupling model (LOPC). The carrier density calculated by using LOPC
fit with experimental data varies from $1.8\times10^{15}$ to $4.2\times10^{17}$
cm$^{-3}$. Room temperature magnetic measurements exhibit ferromagnetic
behavior with non-zero coercivity for all the samples up to 7 T field.
We, for the first time, show the Curie temperature to be above 760
K. Quantitative analysis of magnetic interaction validates the applicability
of Bound Magnetic Polaron Model (BMP) which probably arises from the
exchange interaction of Cr$^{3+}$ ions with related (Si, C) defects.
The polaron density estimated from the BMP fit agrees well with the
carrier density obtained from the line shape fitting of Raman spectra.
\end{abstract}
\maketitle
\begin{singlespace}

\section{INTRODUCTION}
\end{singlespace}

\begin{singlespace}
\noindent For developing a potential spintronics device it is necessary
to have a magnetic semiconductor which is compatible with existing
electronics principles. But to make a semiconductor carrying magnetic
properties is not easy. Inducing magnetic order in a semiconductor
by a transition metal doping has been a trend since Dietl. et al predicted
the room temperature ferromagnetism in wide band gap semiconductor
i.e., Mn (5\%) doped GaN and ZnO \citep{Dietl2000,Reed2001}. But
it turns out that none of these are room temperature ferromagnetic
semiconductors. A recent report on intrinsic ferromagnetic order in
Co doped TiO$_{2}$ has revived interest in dilute magnetic semiconductor
DMS \citep{Saadaoui2016}. It is known that SiC is a potentially wonderful
material for spintronics device on account of its unique properties
under extreme conditions such as high breakdown voltage, low thermal
conductivity and high thermal expansion coefficient and hence can
be used for high temperature and high power electronics applications.
Thus it is interesting to explore the possibility of inducing ferromagnetism
by doping transition metal elements. Among the various transition
metals explored for inducing magnetism in 3C-SiC, the Cr substitution
for Si and its potential application for spintronics device is hardly
explored. We dope Cr in 3C-SiC to check whether it is possible to
induce long range magnetic order in the host matrix via a novel synthesis
technique called carbothermal reduction of silica in rice husk using
plasma reactor \citep{Singh2002,Mishra2011}. We must emphasize that
although there are some reports in the literature on Cr doped 3C-SiC
single crystal, thin film and others, there are no reports on detailed
quantitative analysis of magnetic interactions, and its temperature
dependence \citep{Jin2008,HUANG2007,Yoon2005,Los2009} to the best
of our knowledge. Detailed investigation was carried out to study
the temperature dependence of X-band and Q-band Electron Paramagnetic
resonance both in high as well as low temperature regimes. DC Magnetization,
and temperature variation Raman spectroscopy were carried out to check
the crystalline quality of the sample. Quantitative analysis of magnetization
is done using Bound Magnetic Polaron (BMP) model and role of lattice
defects such as Si/C vacancies along with carrier concentration of
transition metal element is discussed in details. A good agreement
between the theory and experimental data has been achieved while considering
the coupling between Longitudinal Optical (LO) mode with Plasmon induced
by the surface deformation potential due to doping of transition metal
element. So the LO phonon behavior is largely affected by the Longitudinal
Optical Plasmon Coupling (LOPC) present in the system. In this report,
we study the role of carrier concentration and lattice defects in
the system as a function of temperature.
\end{singlespace}

\onecolumngrid

\begin{figure}
[t]
\centering{}\includegraphics[scale=0.3]{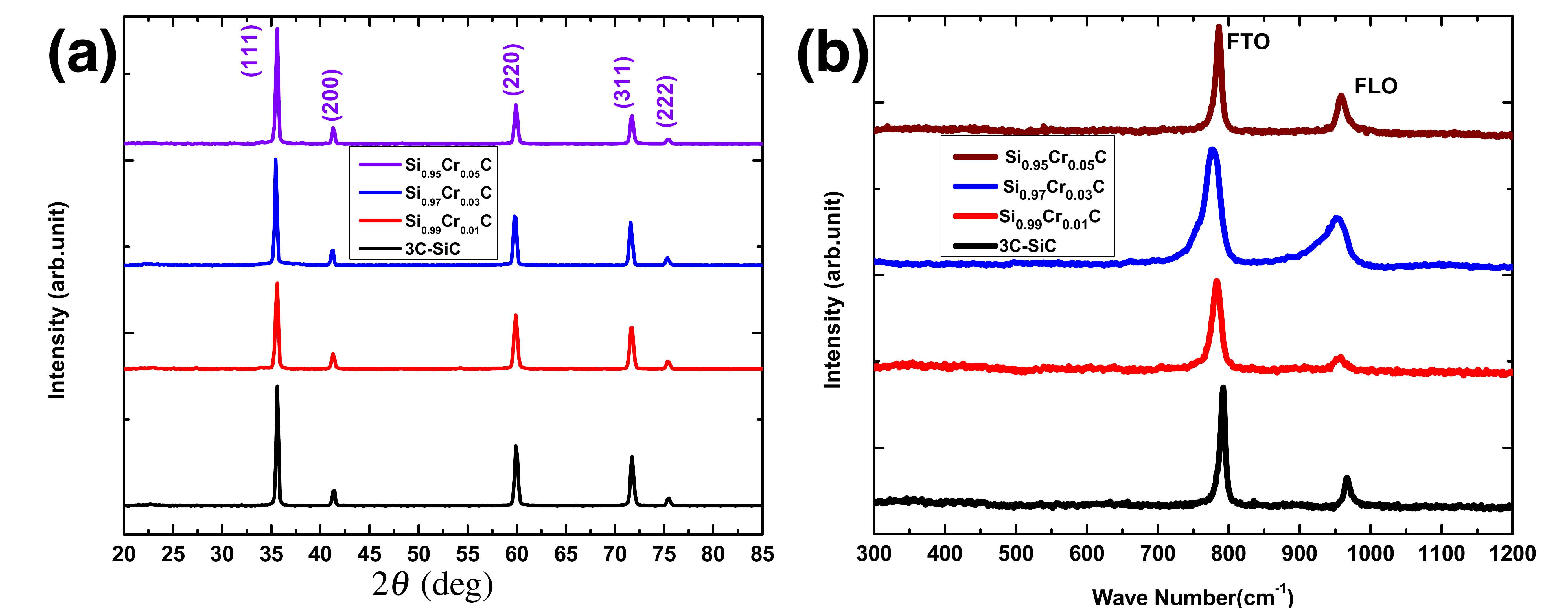}\caption{\label{fig:Raman-spectra-comparison-room-temp}(a) X-Ray Diffraction
pattern of undoped and Cr doped 3C-SiC shows the single phase nature
of the samples. (b) Raman Spectra of undoped and Cr doped 3C-SiC at
room temperature. It shows a clear shift and broadening of the spectra
with increase in Cr concentration in the host matrix. This can be
attributed to the incorporation of Cr in the pristine sample.}
\end{figure}

\begin{figure}
[t]
\begin{centering}
\includegraphics[scale=0.27]{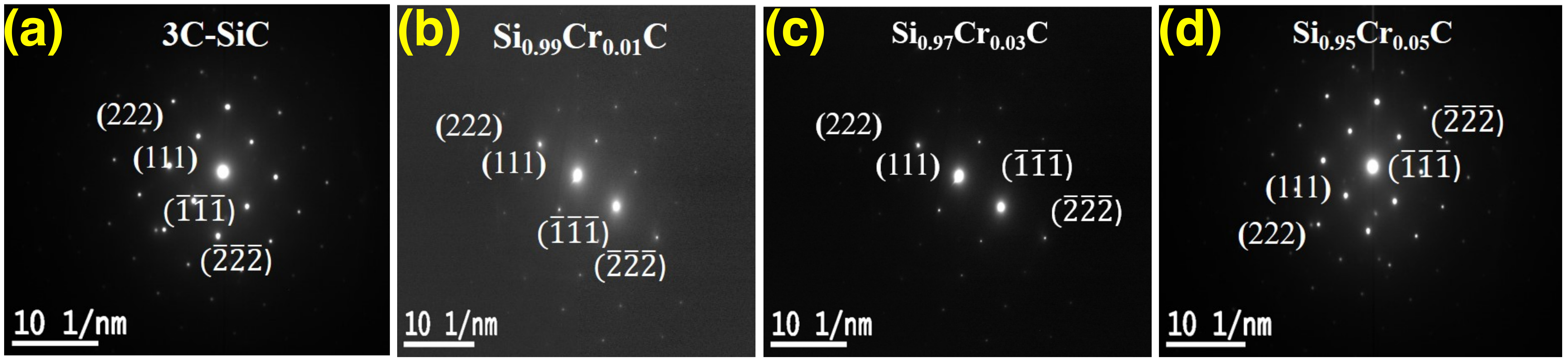}
\par\end{centering}
\caption{\label{fig:Transmission-electron-microscopy SAED pattern of undoped and Cr doped 3C-SiC}Transmission
electron microscopy of undoped and Cr doped 3C-SiC. The figure clearly
shows the four fold symmetry representing the cubic crystal structure
preserved in both undoped and Cr doped 3C-SiC.}
\end{figure}

\twocolumngrid
\begin{singlespace}

\section{Experimental Technique}
\end{singlespace}

\begin{singlespace}
The Cr doped bulk SiC polycrystalline samples were prepared by carbothermal
reduction method. Rice husk was mixed with appropriate amounts of
Cr$_{2}$O$_{3}$. The growth temperature and time were fixed at 1600
°C and 15 mins respectively. The details of synthesis of SiC from
rice husk can be found elsewhere \citep{Singh2002}. Typical experimental
conditions are Argon gas flow -- 2 LPM; current -- 50A, and load
voltage -- 300V. The possible reactions for the formation of SiC
from Rice Husk can be written as
\begin{equation}
\text{SiO}_{2}\text{(amorphous)}+\text{3C}\text{(amorphous)}\to\text{SiC}+\text{2CO},
\end{equation}
where as the chemical reaction for Cr incorporation in SiC is as follows
\[
\text{Cr\ensuremath{_{2}}}\text{O}_{3}+2\text{SiO}_{2}+3\text{C}\to2\text{CrSiC}+\text{C}\text{O}_{2}+5\text{O}\uparrow.
\]
The structural characterization was carried out using Rigaku Smart
lab X-ray diffractometer (XRD) with 9 kW power generator at room temperature
and Horiba Jobin- Yvon HR-800 Micro Raman spectrometer with 488 nm
laser wavelength. The quantum design SQUID-Vibrating Sample Magnetometer
(VSM) was used to measure the magnetization at low temperature\textcolor{red}{{}
}(5K -350K) up to 7 T. The high temperature measurements were carried
out using a Lakeshore VSM. The valence state of unpaired electrons,
dipolar interaction and anisotropy in the system was probed using
electron paramagnetic resonance spectrometer JEOL model JES FA 200.
Fig. \ref{fig:Raman-spectra-comparison-room-temp}(a) shows the X-Ray
diffraction pattern of undoped and Cr doped 3C-SiC. All diffraction
peaks correspond to the cubic phase of SiC with symmetry group F$\bar{4}$3m.
There was no signature of secondary phase of SiC or Cr oxide phase
and Cr composite which indicates that $\text{Cr}^{3+}$ cation replaces
the Si$^{+4}$ cation site in the 3C-SiC. It was found that the diffraction
peaks of Cr doped SiC were shifted to lower angles and intensity was
reduced in addition to an increase in Full Width at Half Maximum (FWHM).
The corresponding expansion of the lattice parameter is expected because
the ionic radius of dopant Cr$^{3+}$ ion (0.52 Å) was larger than
that of the host Si$^{4+}$ (0.40 Å).
\end{singlespace}

The average crystallite size of the Cr doped SiC was (determined by
using Scherror\textquoteright s formula) found to be $<100$ nm. The
Raman spectra of undoped and Cr doped 3C-SiC is  shown in the Fig.
\ref{fig:Raman-spectra-comparison-room-temp}(b). It clearly shows
the variation of intensity and peak position with Cr concentrations.
In order to get a better understanding of the systems, we use LOPC
model to fit the experimental data (Temperature variation Raman spectra)
and give the explanation for non monotonous variation of peak position
and intensity of Raman spectra as a function of temper-

\newpage
\onecolumngrid

\begin{figure}
[t]
\begin{centering}
\includegraphics[scale=0.2]{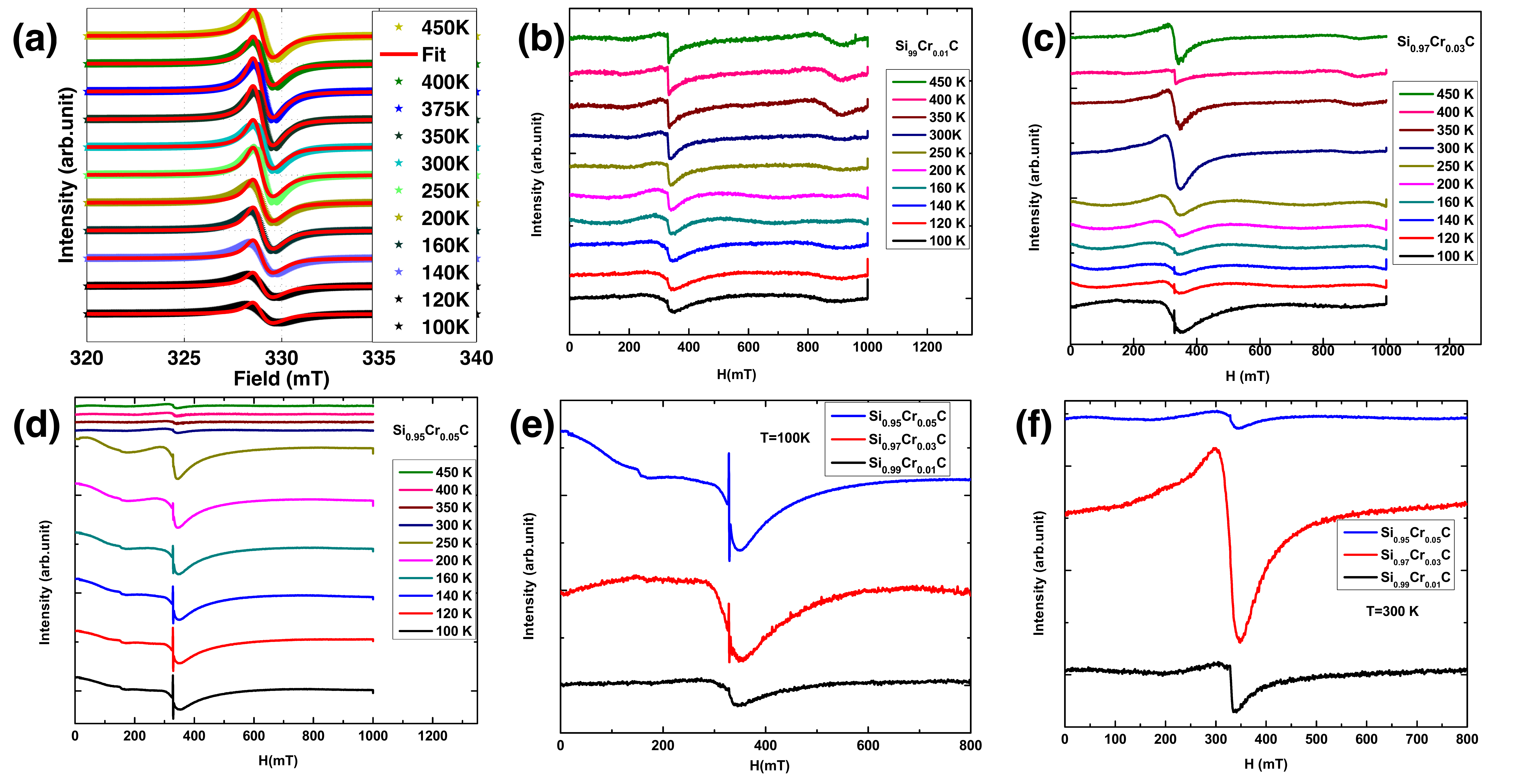}
\par\end{centering}
\caption{\label{fig:Temperature-Variation-of-X-Band EPR spectra }Temperature
Variation of X-band Electron paramagnetic resonance of undoped and
Cr doped 3C-SiC. Fig. (a) shows the temperature variation of EPR spectra
of 3C-SiC. It clearly indicates an increase in intensity with an increase
in temperature. Fig. (b, c, d) show Cr doped 3C-SiC. A sharp line
along with a forbidden transition is an artifact of increase in Cr
concentration. Fig. (e) shows a sharp line present in the spectrum
which goes on decreasing with decrease in Cr concentration. In addition
to this, there is a forbidden transition appearing in case of excess
Cr doping in the system. Fig. (f) represents the effect of Cr concentration
on EPR spectra at 300K. It shows that 3\% Cr doped sample has more
line width and intensity as compared to other two Cr (1\%, \& 5\%). }
\end{figure}

\twocolumngrid

\noindent ature and concentrations. Transmission electron microscopy
selected area diffraction pattern (SAED) is shown in Fig. \ref{fig:Transmission-electron-microscopy SAED pattern of undoped and Cr doped 3C-SiC}
which further confirms the crystalline quality of the samples.
\begin{singlespace}

\section{Electron Paramagnetic Resonance}
\end{singlespace}
\begin{singlespace}

\subsection{X Band EPR Spectra of Cr Doped 3C-SiC}
\end{singlespace}

\begin{singlespace}
Electron Paramagnetic Resonance (EPR) measurements were carried out
using a JEOL X-band (frequency - 9.5 GHz) spectrometer using a rectangular
cavity with 100 kHz field modulation and phase sensitive detection.
The (EPR) absorption peaks for pure 3C-SiC are located around $329.05$
mT, corresponding to $g=2.00$ with the line width as narrow as 0.98
mT. This confirms the presence of vacancies in the host matrix \citep{Bouziane2008,Son2006}
as shown in the Fig. \ref{fig:Temperature-Variation-of-X-Band EPR spectra }(a).
In order to probe the nature of magnetic phase, spin dynamics, spin
relaxation and internal field, the temperature variation of EPR is
an effective tool. Fig. \ref{fig:Temperature-Variation-of-X-Band EPR spectra }(a,
b, c, d) shows the EPR spectra of doped and undoped 3C-SiC with Cr
(1, 3, 5) \% concentrations recorded at different temperatures (100K
- 450K) using X-band rectangular cavity. The spectrum of pure 3C-SiC
sample has shown the feature of resonance at resonance field Hr =
$329.05$ mT, corresponding to $g=2.00$ which could be due to the
defects present in the sample ($V_{\text{Si}}$, $V_{c}$) \citep{Liu2011}.
The line shape parameters and $g$ value can be estimated from the
analysis of the EPR spectra by fitting it to the two component Lorentzian
function \citep{Yang2011,Janhavi2004,Ivanshin2000}. It comprises
of two circular components of the exciting linearly polarized microwave
magnetic field. It can be written as follows
\end{singlespace}

\begin{equation}
\frac{dP}{dH}\mathrm{=}\frac{2A}{\pi}\frac{d}{dH}\left(\frac{\Delta H}{4(H-H_{0})^{2}+\text{\ensuremath{\Delta H^{2}}}}+\frac{\Delta H}{4(H+H_{0})^{2}+\text{\ensuremath{\Delta H^{2}}}}\right)
\end{equation}
where $A$ represent area of the absorption curve, $H_{0}$ is the
Resonance field and $\Delta H$ is the full width at half maximum
in the absorption curve. 

\begin{figure}
[t]
\begin{centering}
\includegraphics[scale=0.3]{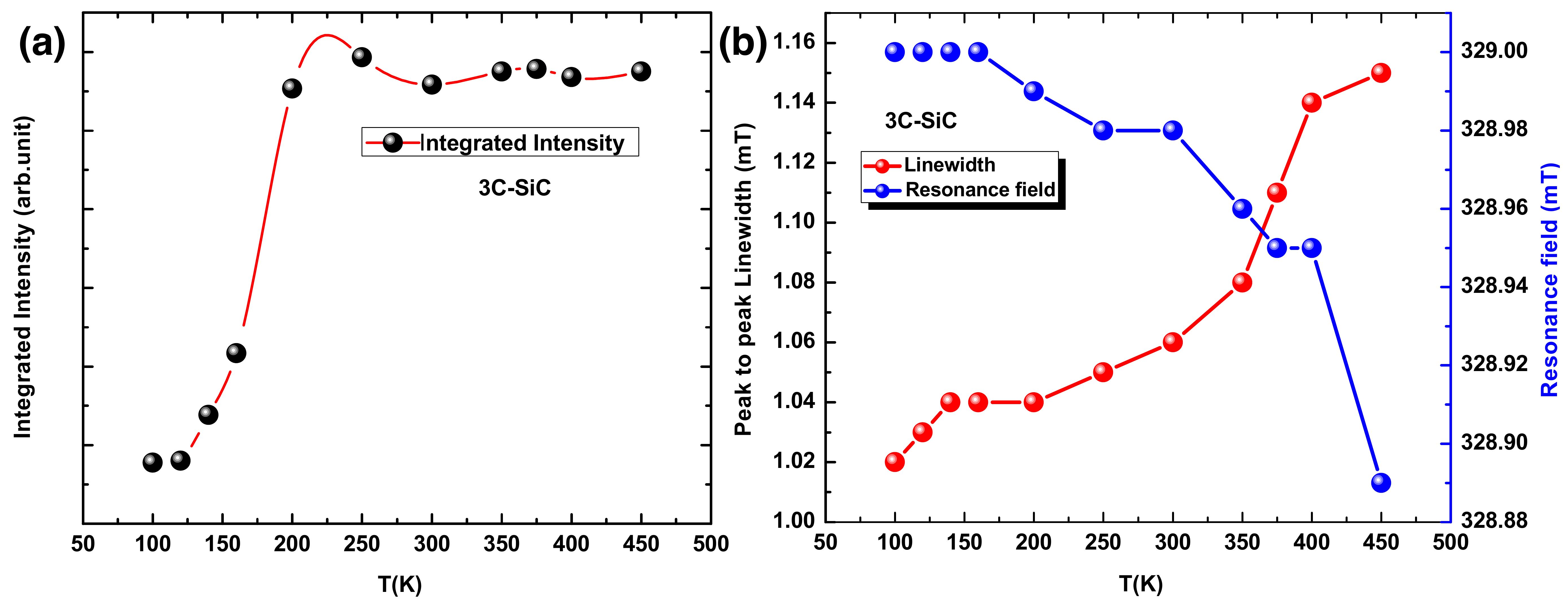}
\par\end{centering}
\caption{\label{fig:Variation-of-Integrated-Intensity}(a) Variation of Integrated
Intensity with temperature of 3C-SiC (b) peak to peak line width and
variation of resonance field with temperature for 3C-SiC.}
\end{figure}

The pick width can be obtained from the fitting by using the formula
$\Delta H$$_{pp}=\Delta H/$$\sqrt{3}$, and $g$ value can be calculated
from the fit parameter $H_{0}$ by resonance condition $g=h\nu/\mu_{B}H_{0}$,
where $h$ is Planck constant, $\nu$ is the microwave frequency and
$\mu_{B}$ is the Bohr magneton. The variation of integrated intensity,
Resonance field and peak to peak line width is shown in the Fig \ref{fig:Variation-of-Integrated-Intensity}.
Here the temperature dependence of line width can be explained by
sum of both dipolar interaction and exchange narrowing mechanism in
3C-SiC. The ferromagnetic signature present in the sample can be attributed
to the exchange interaction present in the system \citep{Peng2018,Zorko2014,Zorko2015,Zimmermann2016}. 

Even at very low doping concentration (1\% Cr) in 3C-SiC, the EPR
signal mainly shows very broad line. The reduction

\begin{singlespace}
\newpage
\onecolumngrid 
\end{singlespace}

\twocolumngrid

\noindent in both line width and peak intensity has been observed
with an increase in Cr$^{3+}$concentration. This broadening can be
attributed to dipolar interaction between isolated Cr$^{3+}$ ions
at low concentration. The anti ferromagnetic coupling due to exchange
interaction between Cr$^{3+}$nearby ions which is responsible for
the reduction in intensity with an increase in Cr$^{3+}$concentration
in host matrix has been observed in the EPR spectra both at 100 K
as well as 300 K shown in Fig. \ref{fig:Temperature-Variation-of-X-Band EPR spectra }(e,
f). Signal-I is a sharp line corresponding to unresolved six hyperfine
lines. However, the forbidden lines ($\Delta M_{I}=\pm1$) present
in the signal I are due to slight lattice distortion from an ideal
tetrahedral structure. In Fig \ref{fig:Temperature-Variation-of-X-Band EPR spectra }(b)
is no forbidden transition is observed in the EPR spectra for Si$_{0.99}$Cr$_{0.01}$C,
instead a small sharp line observed at 100K disappear with an increase
in temperature. Another broad resonance is also obtained which increases
in intensity with temperature. The spectrum of (Si$_{0.97}$Cr$_{0.03}$C
) is shown in Fig. \ref{fig:Temperature-Variation-of-X-Band EPR spectra }(c)
also exhibits 3 different signals, such as 
\begin{enumerate}
\begin{singlespace}
\item A dipolar interaction giving rise to a spectrum whose intensity increases
and line width decreases with increase in temperature
\item A low field forbidden transition because of local distortion due to
replacement of Cr$^{3+}$ ion in place of Si$^{4+}$
\item A sharp line corresponding to Cr$^{3+}$ ion available on the surface
of the sample, whose intensity decreases with increase in temperature. 
\end{singlespace}
\end{enumerate}
\begin{singlespace}
Si$_{0.95}$Cr$_{0.05}$C EPR spectra shown in Fig. \ref{fig:Temperature-Variation-of-X-Band EPR spectra }(d)
indicates the dipolar interaction. It is possible that hyperfine average
and demagnetization field is responsible for broad line observed in
the sample. This dipolar broadening may be due to an interaction between
nearest neighbor Cr$^{3+}$-- Cr$^{3+}$ ions at higher concentration
dopants in the host matrix. There is low field resonance for all the
temperatures. This could be due to a forbidden transition caused by
a local distortion from regular tetrahedron by substitution of Cr$^{3+}$
in place of Si$^{4+}$. With an increase in temperature from 100 K
to 450 K, Signal (I) decreases in intensity and finally disappears.
This is due to the presence of Cr$^{3+}$ ions on the surface. Again,
signal (II) intensity increases with increasing temperature and suddenly
falls drastically at 300 K. This might be due to a phase transition
(magnetic) in the system as is evident from Fig. \ref{fig:Temperature-Variation-of-X-Band EPR spectra }(e,
f). The $g$ value for 1\%, 3\%, and 5\% Cr doped 3C-SiC increases
with an increase in composition due to an increase in dipolar interaction.
But at 5\% Cr substitution, more Cr$^{3+}$ sits at the surface and
dipole-dipole interaction intensity decreases as compared to the sharp
line produced by Cr$^{3+}$ ions. So the Cr$^{3+}$ ion doping, affects
the magnetic property by changing the structure of 3C-SiC with an
increase in tetragonal phase of this compound and influences the Si/C
vacancies defect formation process in the system. This is the reason
for the formation and destruction of FM order in Cr doped 3C-SiC.
There appears a ferromagnetic resonance (FMR) line due to Si/C vacancies
(defects) in Cr doped samples along with Cr$^{3+}$ lines. Further,
there appears an intense and wide EPR line due to the interaction
among the Cr$^{3+}$ ions in the system on account of excess doping.
The intensity and width of the line increases with increase in concentration
\citep{Sushil2009,Kusuma2018}.
\end{singlespace}
\begin{singlespace}

\subsection{Q Band EPR Spectra of Cr Doped 3C-SiC}
\end{singlespace}

\begin{figure}
[th]
\begin{centering}
\includegraphics[scale=0.2]{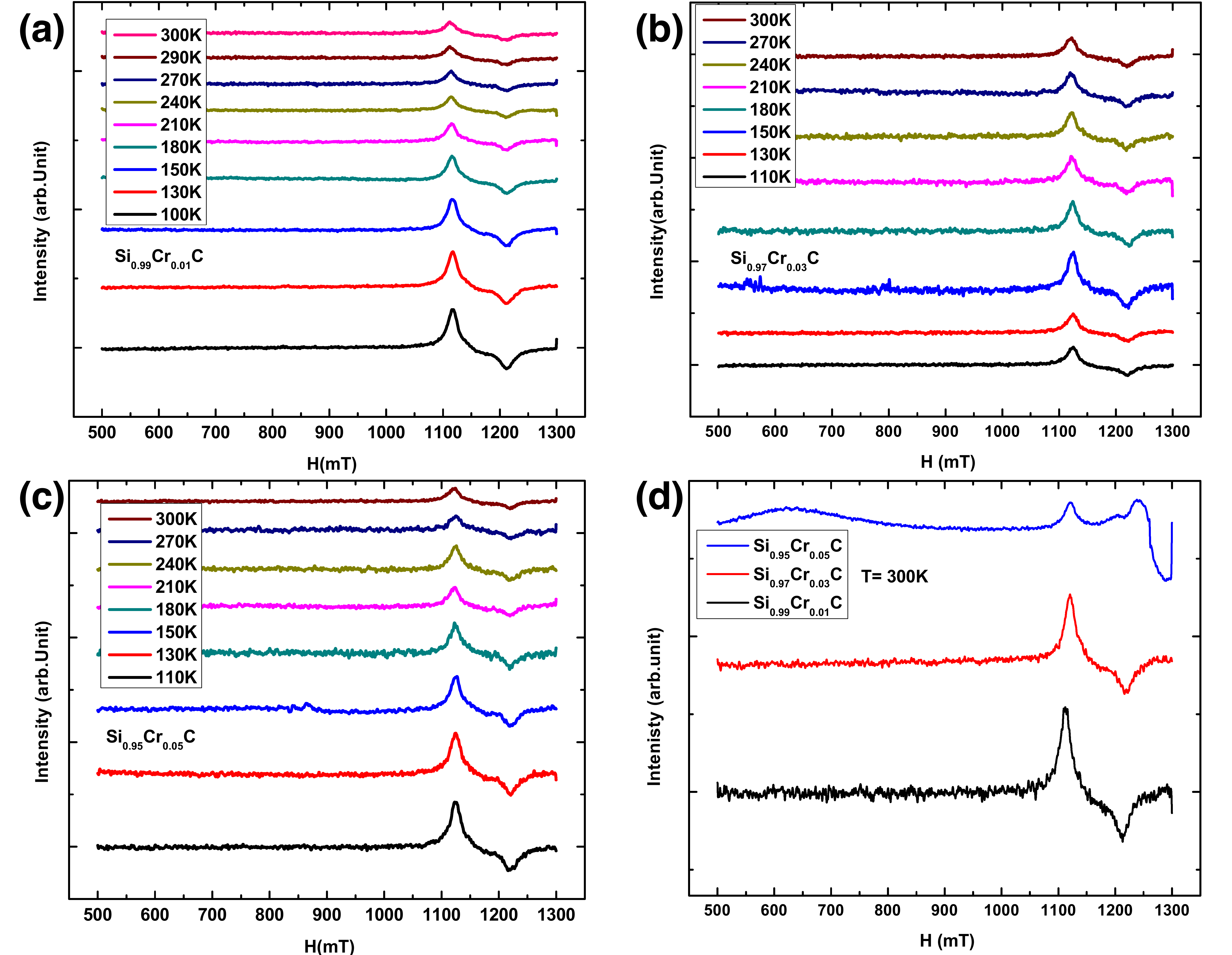}
\par\end{centering}
\caption{\label{fig:Q-band-Electron-Spin Resonance in undoped and Cr doped 3C-SiC}
Q-band-EPR Spectra of undoped and Cr doped 3C-SiC. Fig. (a, b, c)
represent the Cr$^{3+}$ ion transition in Cr doped 3C-SiC. Fig. (d)
represents all three transitions due to excess doping of Cr$^{3+}$
for both low and high fields have been shown. Moreover, there is an
additional peak which corresponds to the defect present in the system.}
\end{figure}

\begin{singlespace}
We have recorded the EPR spectra of 3C-SiC with chromium concentrations
1\%, 3\%, and 5\% in Q-band (35.56 GHz) at different temperatures.
The spectra are shown in Fig. \ref{fig:Q-band-Electron-Spin Resonance in undoped and Cr doped 3C-SiC}(a,
b, c, d,) where Cr$^{3+}$ has d$^{3}$ electronic configuration with
three unpaired electrons and normally is a quartet ($S=3/2$) ground
state. Depending on the symmetry of the Cr$^{3+}$ site, and the distribution
of charge compensating \textquoteleft holes\textquoteright{} around
\end{singlespace}

\newpage
\onecolumngrid

\twocolumngrid

\noindent the defect, the zero-field spin-spin interactions between
the electrons (even in the absence of the external field) lead to
the presence of two Kramer\textquoteright s doublets with spin quantum
numbers $\left|\pm1/2\right\rangle $ and $\left|\pm3/2\right\rangle $
with a separation of $2D$. Here $D$ is the zero-field splitting
when the symmetry of the crystal field is tetragonal. But due to charge
compensating effect in a lattice, where Cr$^{3+}$ is occupying Si$^{4+}$
lattice sites or even in an interstitial site, the crystal field symmetry
may well be rhombic leading to an E-term. The EPR spin Hamiltonian
of Cr$^{3+}$ can be written as 
\begin{equation}
H=D\left[S_{Z}^{2}-\frac{1}{3}S\left(S+1\right)+\frac{E}{D}\left(S_{x}^{2}-S_{y}^{2}\right)\right]+g_{0}\beta SH,
\end{equation}
where $D$ $\rightarrow$ Zero field splitting parameter (ZFS), E
$\rightarrow$ Rhombic splitting parameter at zero field, $E/D$ is
Rhombicity which is 0 (minimum) for an axial system and 0.5 (maximum)
for a rhombic system. $g_{0}\beta SH\rightarrow$ the paramagnetic
contribution to the spin Hamiltonian arising from the isolated electrons.

\begin{figure}
[h]
\centering{}\includegraphics[scale=0.3]{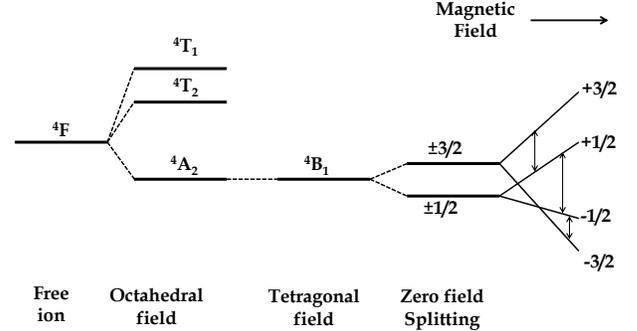}\caption{\label{fig:Energy-Level-Splitting}Energy Level Splitting in the presence
of Magnetic Field}
\end{figure}

\begin{singlespace}
For 3 unpaired electrons ($S=3/2$): The four magnetic spin quantum
numbers (sub levels) for $m_{s}$, can be written as $S_{z}=-3/2$,
-1/2, 1/2, and 3/2. So the energy for the $\pm3/2$ level will be:
$D[9/4-1/3(3/2\times5/2)]=D[9/4-5/4]=D$. And the energy for the $\pm1/2$
level will be: $-D$. Energy level diagram of $S=3/2$, Cr$^{3+}$
progressively under crystal field, zero-field interaction and magnetic
field is shown in Fig. \ref{fig:Energy-Level-Splitting}. The separation
between the two doublet levels is 2D. Typical splitting of the spin
energy levels in the case of $S=3/2$ is given in the above figure.
In the above figure, the three $M_{s}=\pm1$ allowed transitions are
indicated by the double-headed arrows. 
 If the zero-field splitting is very large then 2D could be larger
than the Micro Wave quantum. 
\end{singlespace}

Only two of the allowed transitions, namely, $\left|-3/2\right\rangle $
$\leftrightarrow$ $\left|-1/2\right\rangle $ and $\left|-1/2\right\rangle $
$\leftrightarrow$ $\left|+1/2\right\rangle $ can be seen in the
spectrum, and this is exactly the case in our Cr$^{3+}$ doped 3C-SiC
at X-band at Cr$^{3+}$ concentrations of chromium namely 1\%, and
3\% \citep{Ferretti2004,weil1994}. As the concentration of chromium
is increased from 1\% to 5\%, there is a progressive anti-ferromagnetic
exchange coupling between neighboring chromium ions leading to a reduction
in the magnetization, and also narrowing of the spectral lines. At
5\% concentration of chromium at Q-band frequencies we were able to
see all the three zero-field transitions shown in Fig. \ref{fig:Q-band-Electron-Spin Resonance in undoped and Cr doped 3C-SiC}(d). 

\newpage
\onecolumngrid

\begin{figure}
[ht]
\begin{centering}
\includegraphics[scale=0.2]{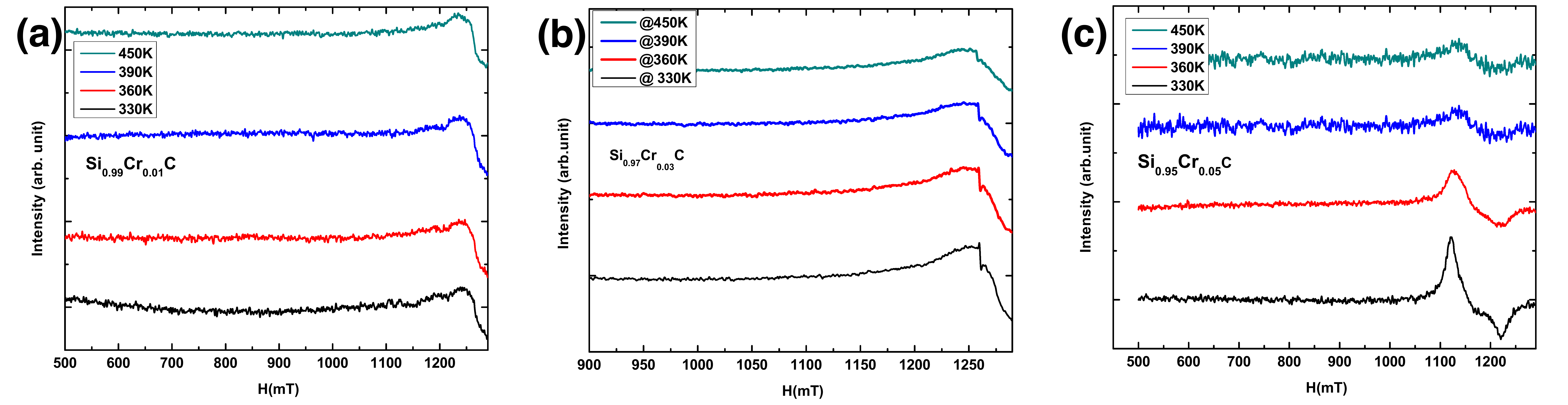}
\par\end{centering}
\caption{\label{fig:Q-band EPR variation with temperature.(330-450)} (a, b,
c) Q-band EPR spectra variation with temperature for (1\%, 3\% \&
5\%) Cr doped 3C-SiC. It shows the Cr$^{3+}$transitions at higher
temperature with different chromium concentration present in the system.}
\end{figure}

\begin{figure}
[ht]
\begin{centering}
\includegraphics[scale=0.2]{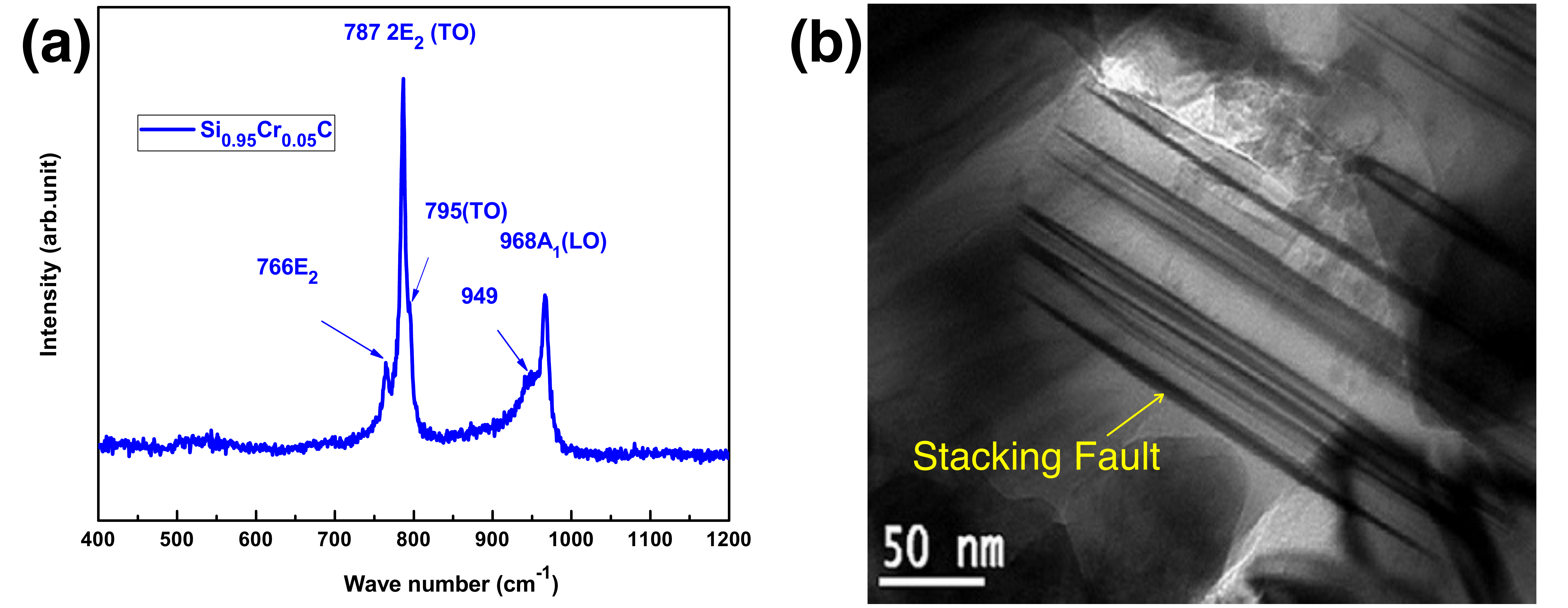}
\par\end{centering}
\caption{\label{fig:Raman spectra of Si0.95Cr0.05C and HRTEM image of stacking faults}(a)
Raman spectra of Si$_{0.95}$Cr$_{0.05}$C and (b) HRTEM image of
stacking faults. Dark lines represent stacking faults formed during
the synthesis of the sample and corresponding peaks are shown in the
Raman spectra.}
\end{figure}

\twocolumngrid

\noindent In systems where the zero-field interaction is large, we
treat each Kramer\textquoteright s doublet as a \textquoteleft fictitious\textquoteright{}
spin $1/2$ system, although the total spin remains $3/2$. Under
this circumstance, the $g$-values arising from these doublets will
occur at $g$-factors which are far removed from $g=2$. For transitions
between $M_{s}=\left|\pm3/2\right\rangle $, $\left|\pm5/2\right\rangle $
and $\left|\pm7/2\right\rangle $, Kramer's doublet will respectively
occur at $g=6$, $10$ and $14$. It is seen in the Q-band spectrum
of 5\% Cr$^{3+}$ doped 3C--SiC that a broad absorption occurs at
$g=4$ corresponding to an allowed transition from $M_{s}=\left|+1/2\right\rangle \leftrightarrow\left|+3/2\right\rangle $
which was not accessible for the X-band quantum, but the Q-band quantum
is adequate to cover this transition. Fig. \ref{fig:Q-band-Electron-Spin Resonance in undoped and Cr doped 3C-SiC}(a)
shows the room temperature Q band ESR spectra of 1\%, 3\% and 5\%
Cr doped 3C-SiC. At 5\% Cr concentration level, we were able to see
all the 3 transitions. The increase in Cr concentration enhances the
formation of defects as confirmed by Raman and HRTEM studies. Moreover,
as the temperature rises, the intensity of Q--Band EPR spectra decreases
with an increase in concentration. This might be due to the fact that
an increase in Cr concentration favors to build up the anti-ferromagnetic
interactions in the system. This is also evident from the magnetic
measurement data. The magnetic moment suffers a reduction as the Cr
concentration increases in the system. There is yet another possibility
in the case of Cr$^{3+}$ concentration exceeding 3\%. 

With an increase in the hole concentration due to charge compensation,
it is possible that excess chromium may act as a hole-trap, generating
a small amount of Cr$^{4+}$ which will get stabilized because of
its charge equivalence to the lattice Si$^{4+}$substitution site,
further favored by its reduction in ionic radius. If this is in fact
what happens upon excess doping of Cr$^{3+}$, then we have a small
percentage of d$^{2}$ in Cr$^{4+}$ which is a triplet state and
is known to have a very large zero-field splitting as observed by
Baranov et. al. \citep{Baranov1994}. These authors report resonances
with $g=3.7$ and $1.77$ for Cr$^{4+}$ in chromium doped 6H-SiC.
It can be seen in our Q-band spectra that at high Cr concentration,
there is a reduction in Cr$^{3+}$ features and simultaneous appearance
of new features at $g=3.8$ and $1.8$. This presumably is due to
the formation of Cr$^{4+}$ by hole trapping on excess Cr$^{3+}$.
In addition to this, Q-band EPR spectra at high temperature (330-450)K
confirms the Cr valence state as Cr$^{3+}$ and magnetic phase transition
in Cr doped 3C-SiC, as shown in Fig. \ref{fig:Q-band EPR variation with temperature.(330-450)}(a,
b, c).
\begin{singlespace}

\section{Raman Study}
\end{singlespace}

\begin{singlespace}
\noindent Raman scattering spectra of 3C--SiC and Cr doped 3C--SiC
with different doping concentrations (1, 3, 5) \% were measured from
100 K to 840 K, see Fig. \ref{fig:Raman-spectra-comparison-room-temp}(b).
It can be seen that there are two characteristics overlapping bands
at 796 cm$^{-1}$ and 789 cm$^{-1}$ observed in the 3C--SiC system
along with some additional defect peaks \citep{Yugami1987,Rohmfeld1998}.
These are assigned to the folded transverse optical (FTO) bands, folded
longitudinal
\end{singlespace}

\newpage
\onecolumngrid

\begin{figure}
[ht]
\begin{centering}
\includegraphics[scale=0.3]{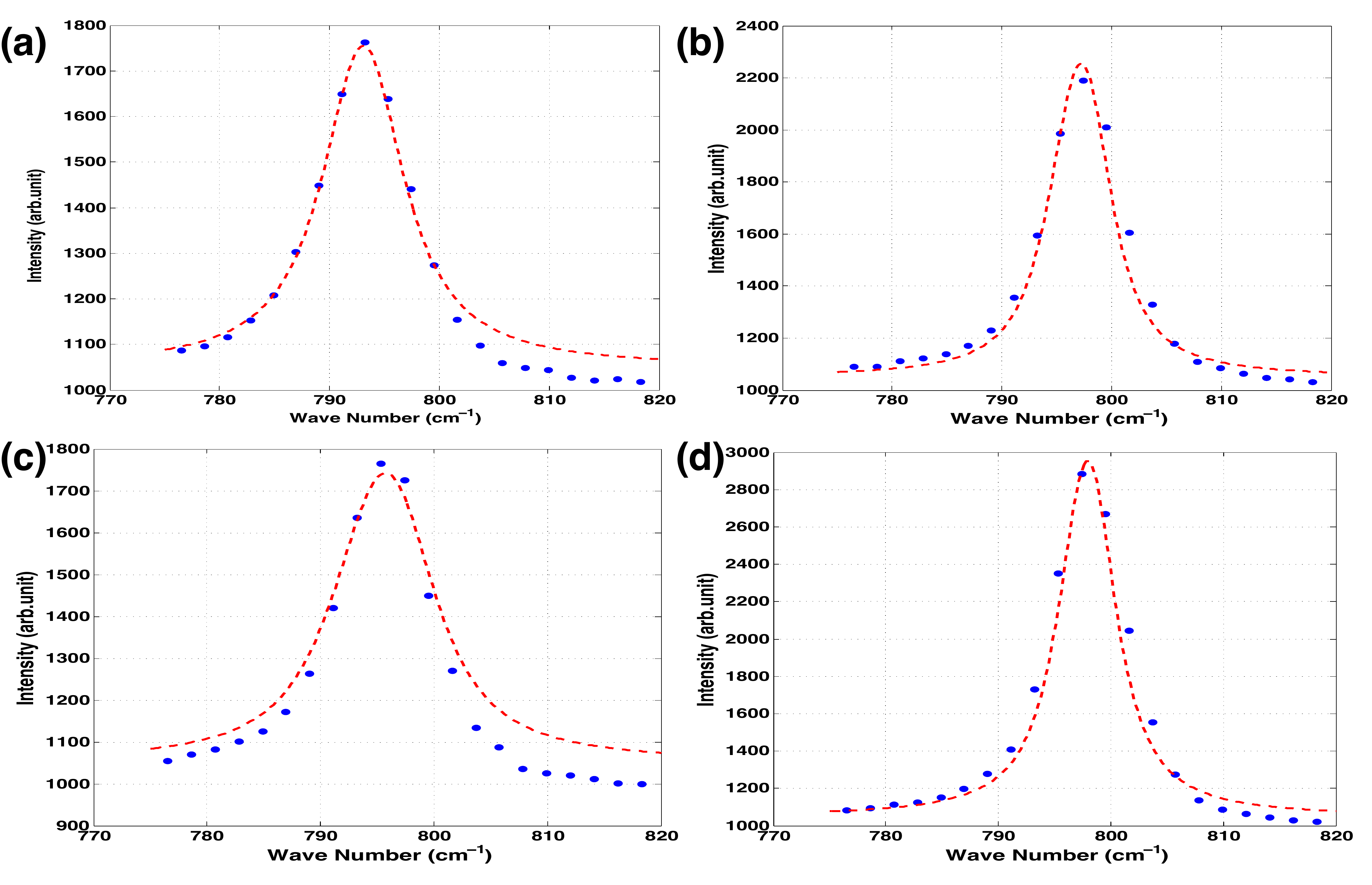}
\par\end{centering}
\caption{\label{fig:TO-Fit-for-undoped and Cr doped 3C-SiC at100K}(a, b, c,
d) TO Fit for undoped and Cr doped 3C-SiC at 100K. In all four cases
blue circles represent the experimental data while dotted lines represent
the fit. }
\end{figure}

\begin{table}
[ht]
\begin{centering}
\begin{tabular}{lcccc}
\hline 
Sample & 3C-SiC & Si$_{0.99}$Cr$_{0.01}$C & Si$_{0.97}$Cr$_{0.03}$C & Si$_{0.95}$Cr$_{0.05}$C\tabularnewline
\hline 
A (cm$^{-1}$) & 795.1 & 795.2 & 795.0 & 800.0\tabularnewline
B (cm$^{-1}$) & 15.0 & 32.5 & 15.0 & 23.7\tabularnewline
L($\mathring{\mathrm{A})}$ & 25.5 & 37.7 & 30.0 & 25.5\tabularnewline
$\Gamma_{0}$(cm$^{-1}$) & 11.7 & 8.5 & 11.7 & 8.5\tabularnewline
\hline 
\end{tabular}
\par\end{centering}
\caption{\label{tab:Fitted-parameters-on-TO-T=00003D100}Fitted parameters
on TO modes for samples, excited by the 488 nm at T =100K.}
\end{table}

\begin{figure}
[t]
\begin{centering}
\includegraphics[scale=0.3]{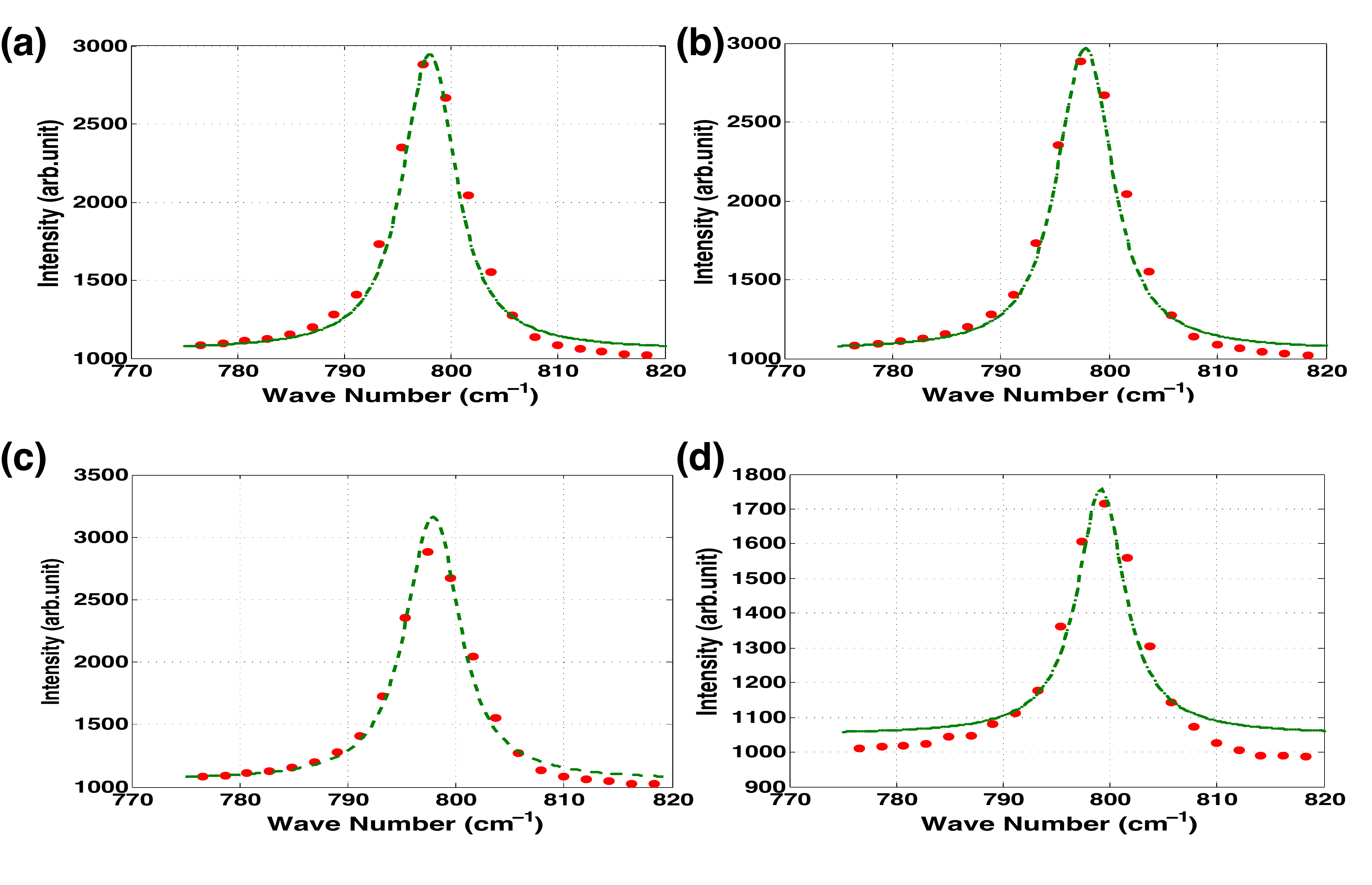}
\par\end{centering}
\caption{\label{fig:TO-fit-of-Raman-Spectra ofundoped and Cr doped sample at 800K}(a,
b, c, d) TO fit of Raman spectra of undoped and Cr doped 3C-SiC at
800K. In all four cases red circles represent the experimental data
whereas dotted lines represent the fit. }
\end{figure}

\twocolumngrid

\noindent optical (FLO) bands, and defect peaks respectively. However
TO band is doubly degenerate. But due to the reduction of symmetry
in the system from cubic to other crystal structure on account of
the formation of other polytypes or defects (stacking faults), the
degeneracy is lifted and transverse component shifts to lower wave
number side. The increase in defect number density broadens the observed
signals and in the case of excess number density it may induce stacking
disorder in the system. Furthermore, signature can be seen in the
spectra as an additional peak at lower wave number as shown in Fig.
\ref{fig:Raman spectra of Si0.95Cr0.05C and HRTEM image of stacking faults}.
As a result of this, unit cell volume of 3C-SiC polytype increases,
and the Brillouin zone is reduced by zone folding phenomena. This
results in the appearance of new phonon modes (i.e. folded modes)
in the spectra \citep{Rohmfeld1998}. A similar effect can also be
realized in a system where the periodicity has been lost due to disorder
or defects present in it. This allows the Raman scattering with wave
vector $\mathbf{k}=\mathbf{0}$ to appear in the spectrum due to the
relaxation in wave vector selection rule. The band at 789 cm$^{-1}$
is a FTO band caused by an effective reduction in symmetry. A separation
of 7 cm$^{-1}$, has been found between TO and FTO bands, similar
to that found in 6H-SiC \citep{Bechelany2007}. Similar to the peak
at (767 cm$^{-1}$), FTO band of the 6H structure, another shoulder
peak at 766 cm$^{-1}$has been identified as a signature of random
stacking fault present in the system. The Eq. \ref{eq:SCM-Model}
used for the fitting of experimental data by using the spatial correlation
model (SCM) is given below:

\begin{singlespace}
\noindent 
\begin{equation}
I(\omega)\propto\int_{0}^{1}\exp\left(\frac{-q^{2}L^{2}}{4}\right)\frac{d^{3}q}{\left[\omega-\omega\left(q\right)\right]^{2}+\left[\Gamma_{0}/2\right]^{2}},\label{eq:SCM-Model}
\end{equation}
here $I(\omega)$ $\rightarrow$ Transverse Optical Raman scattering
intensity, $q\rightarrow$ function of $2\pi/a$, $a$ $\rightarrow$
Lattice constant, $L$ $\rightarrow$ Phonon propagation length characterizing
the quality of the crystals, $\Gamma_{0}\rightarrow$ FWHM of the
Raman peak of TO (LO) phonon of the bulk crystalline 3C-SiC. Fitting
of experimental Raman data with theoretically stimulated spectra for
100K and 800K data are shown respectively in Figs. \ref{fig:TO-Fit-for-undoped and Cr doped 3C-SiC at100K}
and \ref{fig:TO-fit-of-Raman-Spectra ofundoped and Cr doped sample at 800K}.
The corresponding fitting parameters are shown in Tables \ref{tab:Fitted-parameters-on-TO-T=00003D100}
and \ref{tab:Fitted-parameters-on-TO-T=00003D800}. From the analysis
of the TO mode recorded at RT, reduced crystallinity and formation
of stacking faults were observed due to excess doping of Cr$^{3+}$
ions \citep{Rohmfeld1998}. In addition to this, the possibility of
the formation of Longitudinal optical plasmon mode in polar semiconductor
is also more likely. Since SiC is a polar semiconductor, the asymmetric
broadening of FLO mode in Cr doped 3C-SiC can be attributed to the
Longitudinal Optic-
\end{singlespace}

\newpage
\onecolumngrid

\begin{table}
\begin{centering}
\begin{tabular}{lcccc}
\hline 
Sample & 3C-SiC & Si$_{0.99}$Cr$_{0.01}$C & Si$_{0.97}$Cr$_{0.03}$C & Si$_{0.95}$Cr$_{0.05}$C\tabularnewline
\hline 
A (cm$^{-1}$) & 795.0 & 796.0 & 795.0 & 800.0\tabularnewline
B (cm$^{-1}$) & 15.0 & 23.7 & 21.2 & 41.2\tabularnewline
L($\mathring{\mathrm{A})}$ & 25.4 & 35.5 & 20.5 & 20.7\tabularnewline
$\Gamma_{0}$(cm$^{-1}$) & 11.7 & 5.3 & 8.5 & 11.6\tabularnewline
\hline 
\end{tabular}
\par\end{centering}
\caption{\label{tab:Fitted-parameters-on-TO-T=00003D800}Fitted parameters
on TO modes for samples, excited by the 488 nm at T = 800K.}
\end{table}

\begin{figure}
[t]
\begin{centering}
\includegraphics[scale=0.3]{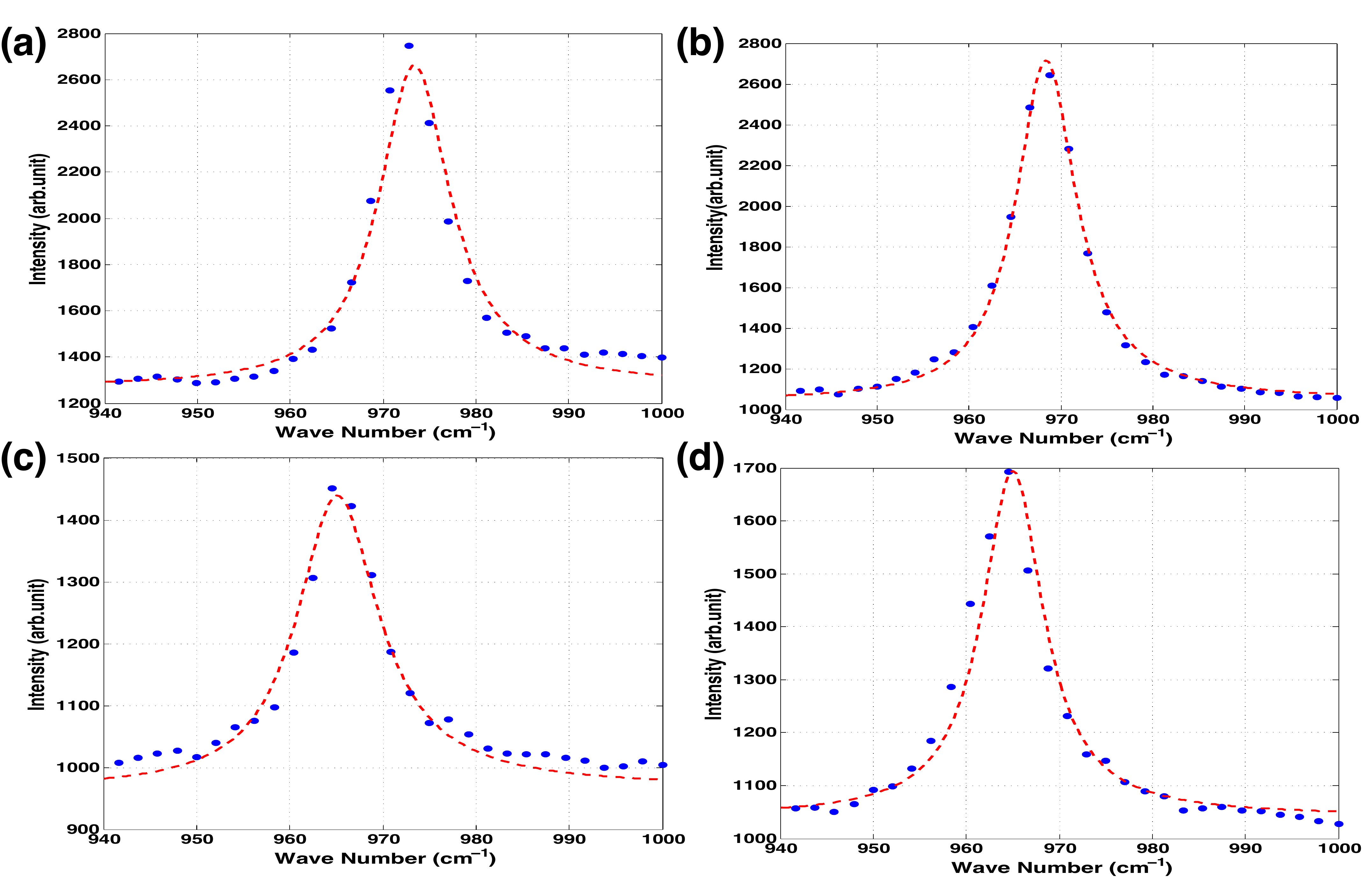}
\par\end{centering}
\caption{\label{fig:LOPC-fit-of-Raman-Spectra-of-undoped-and-Cr-doped at 100K}(a,
b, c, d) LOPC fit of Raman spectra of undoped and Cr doped 3C-SiC
at 100 K. In all four cases blue circles represent the experimental
data while dotted lines represent the fit. }
\end{figure}

\begin{figure}
[ht]
\begin{centering}
\includegraphics[scale=0.3]{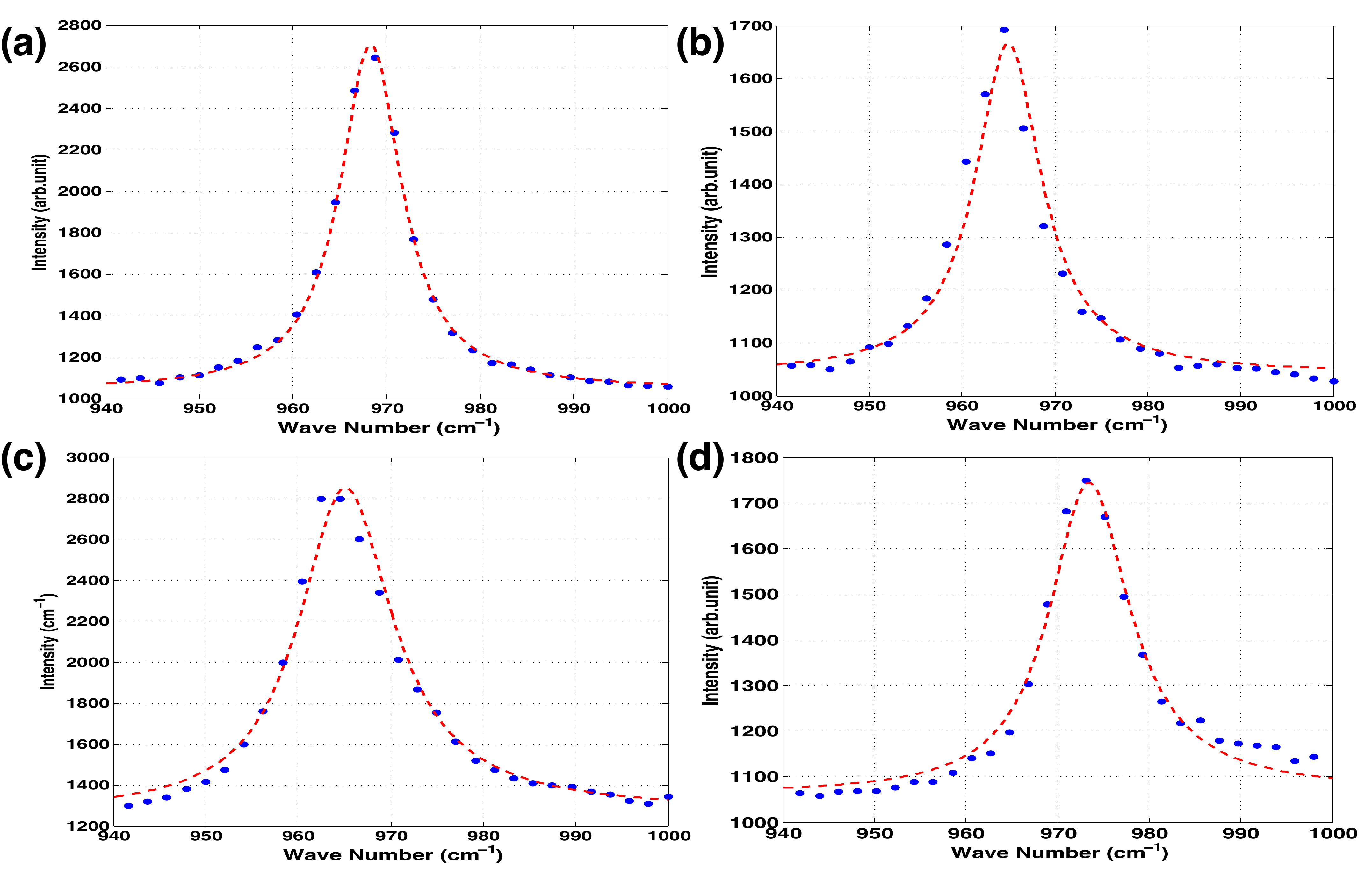}
\par\end{centering}
\caption{\label{fig:LOPC-fit-of-undoped-and-Cr doped 3C-SiC at 800K} (a, b,
c, d) LOPC fit of undoped and Cr doped 3C-SiC at 800K. In all four
cases blue circles represent the experimental data while dotted lines
represent the fit. }
\end{figure}

\twocolumngrid

\noindent al Plasmon Coupling mode (LOPC) present in the system. In
order to check the validity of the assumption, we tried to fit the
experimentally obtained data to LOPC model. It fits quite well as
can be seen in Figs. \ref{fig:LOPC-fit-of-Raman-Spectra-of-undoped-and-Cr-doped at 100K}
and \ref{fig:LOPC-fit-of-undoped-and-Cr doped 3C-SiC at 800K}. Corresponding
fitting parameters can be found in Tables \ref{tab:Fitted-parameters-on-LOPC-T=00003D100}
and \ref{tab:Fitted-parameters-on-LOPC-T=00003D800}. The LOPC fitting
was done with the experimental data in a temperature range (110 --
840) K. There is a linear relationship between carrier concentration
and Raman shift of the LOPC mode in SiC. The non monotonous variation
of LOPC peak with temperature and carrier concentration can be explained
as a coupling between the temperature dependent LOPC mode with carrier
concentration \citep{Liu2010,Sun2013,Chen2017}. The variation of
peak intensity and peak position with temperature and doping concentration
was analyzed by Line shape fitting, using Eq. \ref{eq:I-LOPC-EQN}
and classical dielectric function (CDF) for all Cr doped 3C-SiC samples. 

\begin{singlespace}
\noindent 
\begin{equation}
I_{\text{LOPC}}=\left.\frac{d^{2}\mathbf{S}}{d\omega d\Omega}\right|_{A}=\frac{16\pi hn_{2}}{V_{0}^{2}n_{1}}\frac{\omega_{2}^{4}}{C^{4}}\left(\frac{d\alpha}{dE}\right)\left(\epsilon_{\infty}+1\right)\mathit{A}\text{Im}\left(-\frac{1}{\varepsilon}\right)\label{eq:I-LOPC-EQN}
\end{equation}
where $A$ and $\Delta$ are respectively defined by Eqs. \ref{eq:long_equation}
and \ref{eq:delta}
\end{singlespace}
\begin{widetext}
\begin{singlespace}
\noindent 
\begin{align}
A & =1+2C\frac{\omega_{T}^{2}}{\Delta}\left[\omega_{p}^{2}\gamma(\omega_{T}^{2}-\omega^{2})-\omega^{2}\eta(\omega^{2}+\gamma^{2}-\omega_{p}^{2}\right]+\frac{C^{2}\omega_{T}^{4}}{\Delta\left(\omega_{L}^{2}-\omega_{T}^{2}\right)}\left[\omega_{p}^{2}\gamma\left(\omega_{L}^{2}-\omega_{T}^{2}\right)+\omega_{p}^{2}\eta\left(\omega_{p}^{2}-2\omega^{2}\right)+\omega^{2}\eta\left(\omega^{2}+\gamma^{2}\right)\right]\label{eq:long_equation}
\end{align}
\end{singlespace}
\end{widetext}

\begin{singlespace}
\noindent 
\begin{equation}
\Delta=\omega_{p}^{2}\gamma\left[\left(\omega_{T}^{2}-\omega^{2}\right)^{2}+\left(\omega\eta\right)^{2}\right]+\omega^{2}\eta\left(\omega_{L}^{2}-\omega_{T}^{2}\right)\left(\omega^{2}+\gamma^{2}\right)\label{eq:delta}
\end{equation}

\noindent \noindent In these equations, $\omega_{L}\rightarrow$
LO mode frequency, $\omega_{T}\rightarrow$ TO mode frequency, $\eta$$\rightarrow$
phonon damping constant, $\gamma\rightarrow$ plasma damping constant,
$\alpha\rightarrow$ Polarizability, $E\rightarrow$ Electric field,
$n_{2}$ is the refractive index and $\omega_{2}$ is the scattered
frequency.
\end{singlespace}

\newpage
\onecolumngrid

\begin{table}
\begin{centering}
\begin{tabular}{lcccc}
\hline 
Sample & 3C-SiC & Si$_{0.99}$Cr$_{0.01}$C & Si$_{0.97}$Cr$_{0.03}$C & Si$_{0.95}$Cr$_{0.05}$C\tabularnewline
\hline 
$\omega_{p}$ (cm$^{-1}$) & 5.0 & 48.9 & 92.7 & 136.6\tabularnewline
$\gamma$ (cm$^{-1}$) & 89.4 & 234.4 & 136.5 & 92.78\tabularnewline
$\eta$($\mathring{\mathrm{A})}$ & 1.0 & 2.2 & 1.1 & 1.2\tabularnewline
n $(\times10^{16}$cm$^{-3}$) & 0.1 & 11.4 & 41.0 & 89.1\tabularnewline
\hline 
\end{tabular}
\par\end{centering}
\caption{\label{tab:Fitted-parameters-on-LOPC-T=00003D100}Fitted parameters
on LO- phonon modes for samples, excited by the 488 nm at T =100K.}
\end{table}

\twocolumngrid

\noindent \indent The total Dielectric function due to phonons and
plasmons can be described as

\begin{singlespace}
\noindent 
\begin{equation}
\varepsilon(\omega)=\varepsilon_{\infty}\left(\frac{1+\omega_{L}^{2}-\omega_{T}^{2}}{\omega_{T}^{2}-\omega^{2}-i\omega\eta}-\frac{\omega_{p}^{2}}{\omega\left(\omega+i\eta\right)}\right),
\end{equation}

\noindent 
\begin{equation}
\omega_{p}^{2}=\frac{4\pi ne^{2}}{\varepsilon_{\infty}m^{\ast}},
\end{equation}

\noindent where $C\rightarrow$ Faust-Henry coefficient $=0.35$.
It can also be modified by Cr doping in pure 3C-SiC because of deformed
potential present in the system 

\noindent 
\begin{equation}
C=C_{FH}\left\{ \frac{\varepsilon\left(r\right)_{\infty m}\omega_{T}^{2}\left[\omega\left(r\right)_{\mathrm{Lm}}^{2}-\omega\left(r\right)_{Tm}^{2}\right]}{f_{\infty}\omega_{Tm}^{2}\left(\omega_{L}^{2}-\omega_{T}^{2}\right)}\right\} ^{1/2},
\end{equation}
here $\omega_{p}\rightarrow$ Plasma frequency, $n\rightarrow$ free
carrier concentration, $m^{\star}\rightarrow$ effective mass and
$\varepsilon_{\infty m}\rightarrow$ High field dielectric constant,
$\omega_{Tm}\rightarrow$ Transverse Optical mode frequency, $\omega_{Lm}\rightarrow$
Longitudinal Optical mode frequency. Raman scattering intensity of
LO mode can also be described by a Lorentz profile function for undoped
3C-SiC which can be expressed as

\noindent 
\begin{equation}
I_{A_{1}(\text{LO})}=\frac{I_{0}}{\left(\frac{\omega-\omega_{s}}{\Gamma_{s}}\right)^{2}+1},
\end{equation}
where $\omega_{s}\rightarrow$ LO phonon frequency near the Brillouin
zone center, $\Gamma_{s}\rightarrow$ Line width of peak and $I_{0}\rightarrow$
constant. 
\end{singlespace}

\noindent \indent The LOPC frequency can be calculated by using the
equation

\begin{singlespace}
\noindent 
\begin{equation}
\omega_{\text{LOPC}}^{2}=\frac{\left(\omega_{p}^{2}+\omega_{LO}^{2}\right)+\sqrt{\left(\omega_{p}^{2}+\omega_{LO}^{2}\right)^{2}-4\omega_{p}^{2}\omega_{TO}^{2}}}{2}.
\end{equation}
The carrier density calculated by using LOPC fit with experimental
data varies from 1.8 $\times$10$^{15}$ to 4.2 $\times$10 $^{17}\text{cm}^{-3}$
with temperature. The damping of phonon and plasmon could be the reason
for this behavior in the Raman spectra of Cr doped 3C-SiC. Phonon
life times were calculated using Raman FWHM through the energy-time
uncertainty relation, where $\Delta E$ is the Raman FWHM in units
of cm$^{-1}$ and Planck
\end{singlespace}

\newpage
\onecolumngrid

\begin{table}
[ht]
\begin{centering}
\begin{tabular}{lcccc}
\hline 
Sample & 3C-SiC & Si$_{0.99}$Cr$_{0.01}$C & Si$_{0.97}$Cr$_{0.03}$C & Si$_{0.95}$Cr$_{0.05}$C\tabularnewline
\hline 
$\omega_{p}$ (cm$^{-1}$) & 36.6 & 180.5 & 201.6 & 312.2\tabularnewline
$\gamma$ (cm$^{-1}$) & 90.0 & 136.1 & 168.8 & 224.4\tabularnewline
$\eta$($\mathring{\mathrm{A})}$ & 1.0 & 1.2 & 1.1 & 1.1\tabularnewline
n $(\times10^{16}$cm$^{-3}$) & 6.4 & 155.6 & 194.0 & 465.5\tabularnewline
\hline 
\end{tabular}
\par\end{centering}
\caption{\label{tab:Fitted-parameters-on-LOPC-T=00003D800}Fitted parameters
on LO- phonon modes for samples, excited by the 488 nm at T = 800K.}
\end{table}

\twocolumngrid

\noindent constant $h=\unit[5.3\times10^{-12}]{cm^{-1}s}$. The phonon
lifetime $\tau$ is largely affected by the reduction of the symmetry
present in the system and can be explained by two mechanisms called
(a) phonon-carrier scattering and (b) phonon-phonon scattering respectively.
The life time calculated from Raman spectra lies in the range of picosecond
which indicates that the carrier - phonon interaction scattering is
the dominant mechanism \citep{Peng2016}.
\begin{singlespace}

\section{Isothermal Magnetization Study}
\end{singlespace}

\begin{figure}
[t]
\begin{centering}
\includegraphics[scale=0.2]{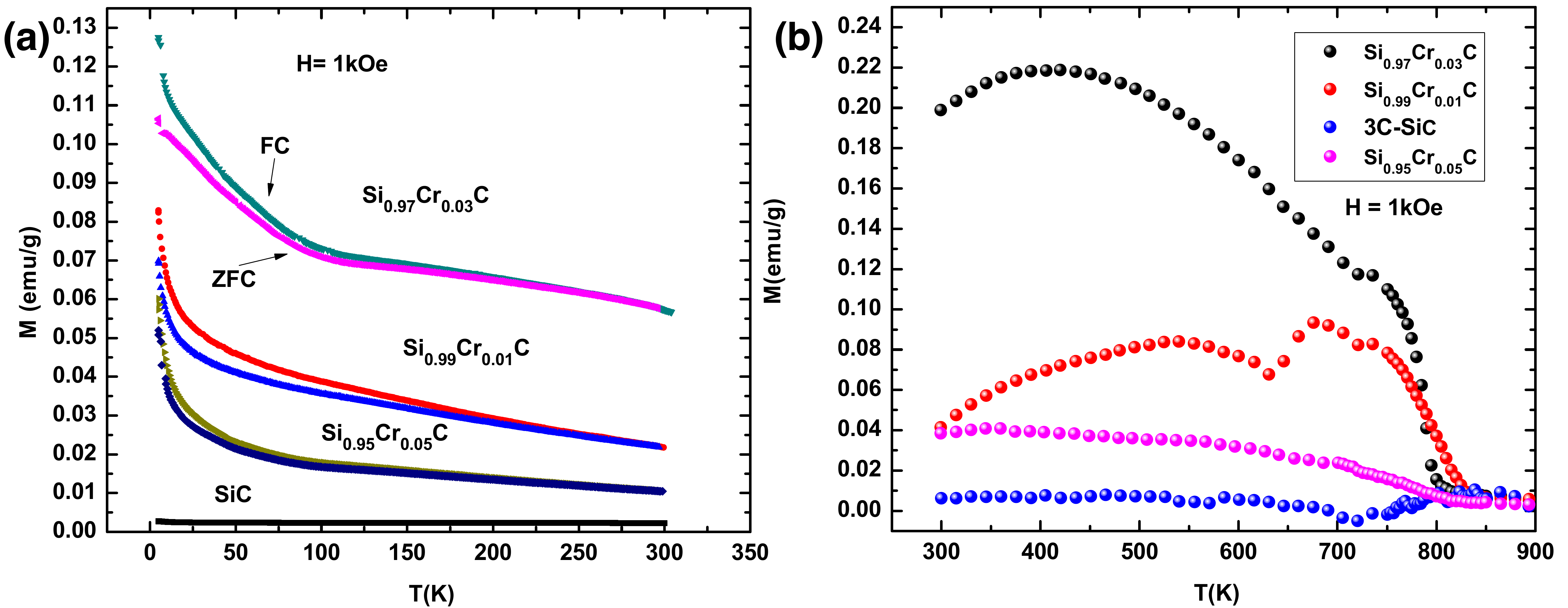}
\par\end{centering}
\centering{}\caption{\label{fig:M-T plots of undoped and Cr doped 3C-SiC}(a) FC-ZFC Plots
of undoped and Cr doped 3C-SiC. The concave upward nature of both
FC-ZFC plots with field indicates the ferromagnetic nature of the
samples. (b) M-T Plots of undoped and Cr doped 3C-SiC samples. It
can be seen that the transition is sharper with increasing Cr concentration
(1\%, 3\%) and becomes broader at 5\% Cr concentration due to the
present of anti ferromagnetic interaction present in the system. }
\end{figure}

\begin{singlespace}
\noindent DC magnetic measurements were carried out using vibrating
sample magnetometer in the low and high temperature regimes. Fig.
\ref{fig:M-T plots of undoped and Cr doped 3C-SiC} shows the FC (field
cool) -- ZFC (zero field cool) plot of undoped and Cr doped 3C-SiC
up to 300K. In the case of pure 3C-SiC, FC-ZFC curves show a small
bifurcation. This might be due to the defects present in the sample.
But the magnitude of magnetization is extremely small. The isothermal
field variation shows small hysteresis with low moment at 5K, whereas
only a diamagnetic response is observed above 80K. The weak ferromagnetic
like response at 5K may be due to the defects (Si or C vacancies)
present in the sample. The electron trapped in such vacant sites (F-Centers)
may interact to give a weak ferromagnetic response. The Cr doped samples
show a clear bifurcation up to room temperature. The clear bifurcation
of FC-ZFC plot of Cr doped sample indicates the ferromagnetic order
present in the system. The Curie temperature ($T_{\text{C}}$) is
found to be above 780 K as shown in the Fig. \ref{fig:M-T plots of undoped and Cr doped 3C-SiC}
and the transition becomes less sharper with an increase in Cr Concentration
\citep{Wang2014}. The Corresponding M-H curve is shown in Fig. \ref{fig:M-H plot comparison of undoped and Cr doped 3C-SiC at 5K ,80K and  350K temperature respectviely.}(a,
b, c). It can be seen that hysteresis goes on decreasing with an increase
in temperature. With increasing Cr concentration, though hysteresis
increases, the saturation magnetization decreases. This may be due
to setting up of anti ferromagnetic order in the system which happens
further on account of excess doping of Cr. This is also evident from
the high temperature M-T plots of all the doped samples. The origin
of long range ferromagnetic order in Cr, Mn, Fe, Co and Ni doped wide
band gap semiconductors have been a topic of research for decades.
As physical properties of materials are intrinsically related to the
structure of the materials, the synthesis procedure, experimental
conditions and morphology plays very decisive role. We have prepared
the Cr doped 3C-SiC by carbothermal reduction of silica from rice 
\end{singlespace}

\newpage
\onecolumngrid

\twocolumngrid

\noindent husk at extremely high temperature ($1500^{\circ}$C). The
experimental conditions is very much liable for the formation of the
lattice defects such as vacancies and interstitial in the system.
The substitution of Cr at the Si site turns out to be more preferable
under these experimental condition. This has been confirmed by structural
and spectroscopic investigation via XRD, Raman and EPR techniques.
Moreover, the Cr substitution at Si sites exhibits more stable ferromagnetism
\citep{Yoon2005}. From EPR, it is confirmed that the Cr ion is present
at both tetrahedral and octahedral co-ordination sites and is in the
trivalent state. The chromium in Cr$^{3+}$state can exhibit low spin
or high spin state. Usually Cr$^{3+}$in low spin state does not induce
magnetism but in high spin state can induce a weak magnetization in
the system. The long rang order in magnetic interaction can be explained
by the exchange interaction between F-Centers (an electron trapped
with positive charge vacancy) in the system. This is known as the
so called Bound Magnetic Polaron Model and was first invoked to explain
the magnetic interaction in Oxide based Dilute magnetic semiconductors
\citep{Kaminski2002,Chiorescu2007,Coey2005}. Here, the {[}Si, C{]}
vacancy, being positively charged, traps the electron and it is referred
to as an F center. Si/C vacancies in this system are adjacent to the
Cr dopants in the lattice, leading to an F-center exchange. In order
to explain the mechanism responsible for magnetic interaction in the
system, the role of magnetic impurity (Carrier concentration) and
vacancy density has to be taken into consideration. In Cr doped 3C-SiC,
the interaction between vacancies (Si/C) nearer to Cr dopants leads
to a ferromagnetic order in the system. It is possible that the doping
of 1\% and 3\% Cr in the host matrix might have increased the (Si/C)
vacancies with the regular substitution at Si sites that resulted
in increased magnetization of the system. Further increase in Cr concentration
might have exceeded the solubility limit. Due to this increase, there
is a net reduction in the magnetization in 5\% Cr doped 3C-SiC. Due
to excess carrier concentration, the distance between Cr--Cr ions
decreases and Cr also goes to a interstitial position in the lattice.
This occupancy destroys the F-centers exchange interaction inducing
anti ferromagnetic interaction the system. The net result is reduction
in total magnetization. This can be confirmed from the sharp line
in EPR spectra shown in Fig. \ref{fig:Temperature-Variation-of-X-Band EPR spectra }(b,
c, d). Thus, along with Cr dopants' density, the F-centers, i.e.,
the Si vacancies play a crucial role in invoking a ferromagnetic exchange
interaction through carrier mediated Polaron interaction. To check
the validity of the BMP model, we have fitted the experimental magnetization
data to the relation \emph{
\begin{equation}
M=M_{0}L\left(x\right)+\chi_{m}H
\end{equation}
\begin{equation}
M_{0}=Nm_{s},\ L\left(x\right)=\coth\left(x\right)-\frac{1}{x},\ X=\frac{m_{\text{eff}}H}{k_{B}T}
\end{equation}
}where

\begin{singlespace}
\noindent $ML_{0}(x)$ $\rightarrow$ BMP contribution 

\noindent $\chi_{m}H$ $\rightarrow$ Paramagnetic contribution 

\noindent $M_{0}=Nm_{s}$ 

\noindent $N$ $\rightarrow$ No of BMP involved in the interaction.

\noindent $m_{s}$ $\rightarrow$ Effective spontaneous moments per
BMP

\noindent $L(x)$ $\rightarrow$ Langevin function 

\noindent $X$ $\rightarrow$ True spontaneous moment per BMP and

\noindent at higher temperature $m_{s}=m_{\text{eff}}$ $M_{0}$,

\noindent $m_{s}$, $m_{\text{eff}}$ $\rightarrow$ Variable in fitting
process

\noindent $m_{\text{eff}}$ $\rightarrow$ The spontaneous moment
per BMP = 10$^{-17}$

\begin{figure}
[t]
\begin{centering}
\includegraphics[scale=0.2]{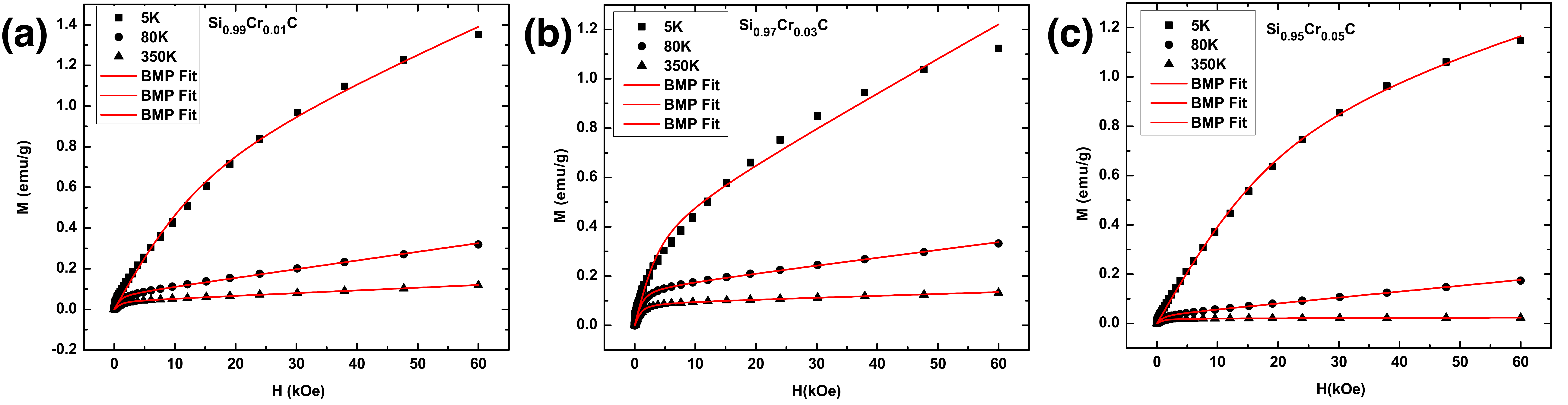}
\par\end{centering}
\caption{\label{fig:.BMP fit with experimental data of Cr doped 3C-SiC} BMP
fit with experimental data of Cr doped 3C-SiC.}
\end{figure}

\noindent \indent Details of the fitting parameters have been given
in Table \ref{tab:BMP fit of Cr doped 3C-SiC}. We propose the BMP
model to explain the high temperature behavior of Cr doped 3C-SiC.
To build up a long range BMP percolation in the system, the concentration
of BMP has to be 10$^{20}$/cm$^{3}$. However, the number density
obtained from the fit is smaller than this number. The fitting of
experimental magnetization data to the theoretical BMP model is shown
in the Fig. \ref{fig:.BMP fit with experimental data of Cr doped 3C-SiC}(a,
b, c). The calculated low concentration of BMPs can\textquoteright t
be held responsible for the observed magnetic interaction in Cr doped
3C-SiC. The effective Bound Magnetic Polaron radius is found to be
$\simeq$ 17 nm. The long range magnetic order in the system can be
established due to BMP percolation in the presence of the lattice
defect in the system. In addition to that, the role of Magnetic impurity
is also significant. If there is more BMP density, the possibility
of the interaction between two isolated Cr ions will be less, otherwise
there will be isolated Cr-Cr ions which can have rare FM interaction
\end{singlespace}

\newpage
\onecolumngrid

\begin{figure}
[t]
\begin{centering}
\includegraphics[scale=0.2]{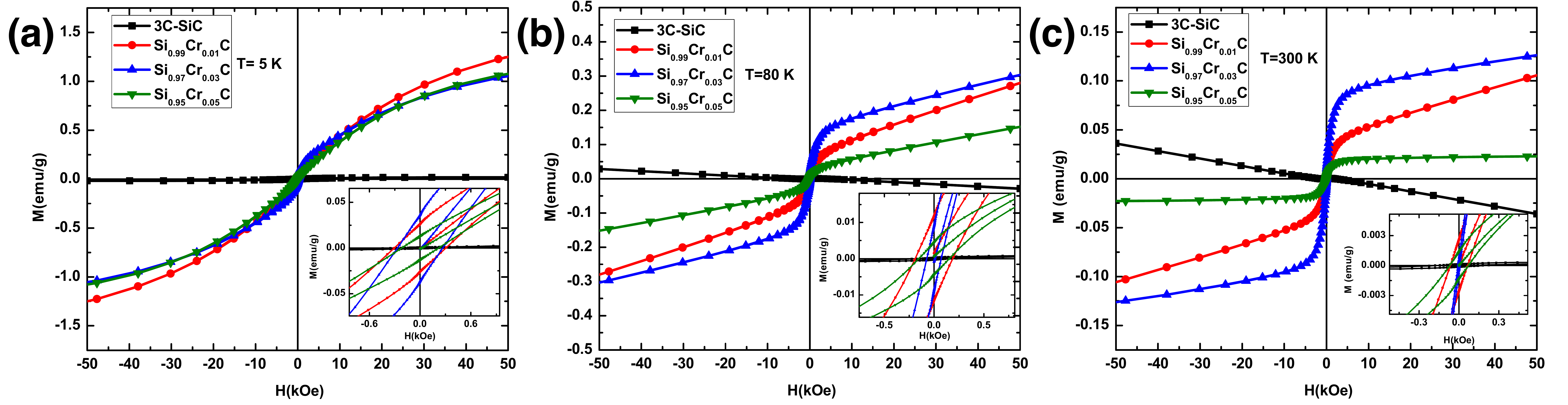}
\par\end{centering}
\centering{}\caption{\label{fig:M-H plot comparison of undoped and Cr doped 3C-SiC at 5K ,80K and  350K temperature respectviely.}(a,
b, c) M-H plots of undoped and Cr doped 3C-SiC at 5K, 80K and 350K
respectively. Pure 3C-SiC shows clear diamagnetic behavior with increase
in temperature. On the other hand, coercivity of Cr doped samples
decreases with increase in temperature. }
\end{figure}

\begin{figure}
[t]
\begin{centering}
\includegraphics[scale=0.2]{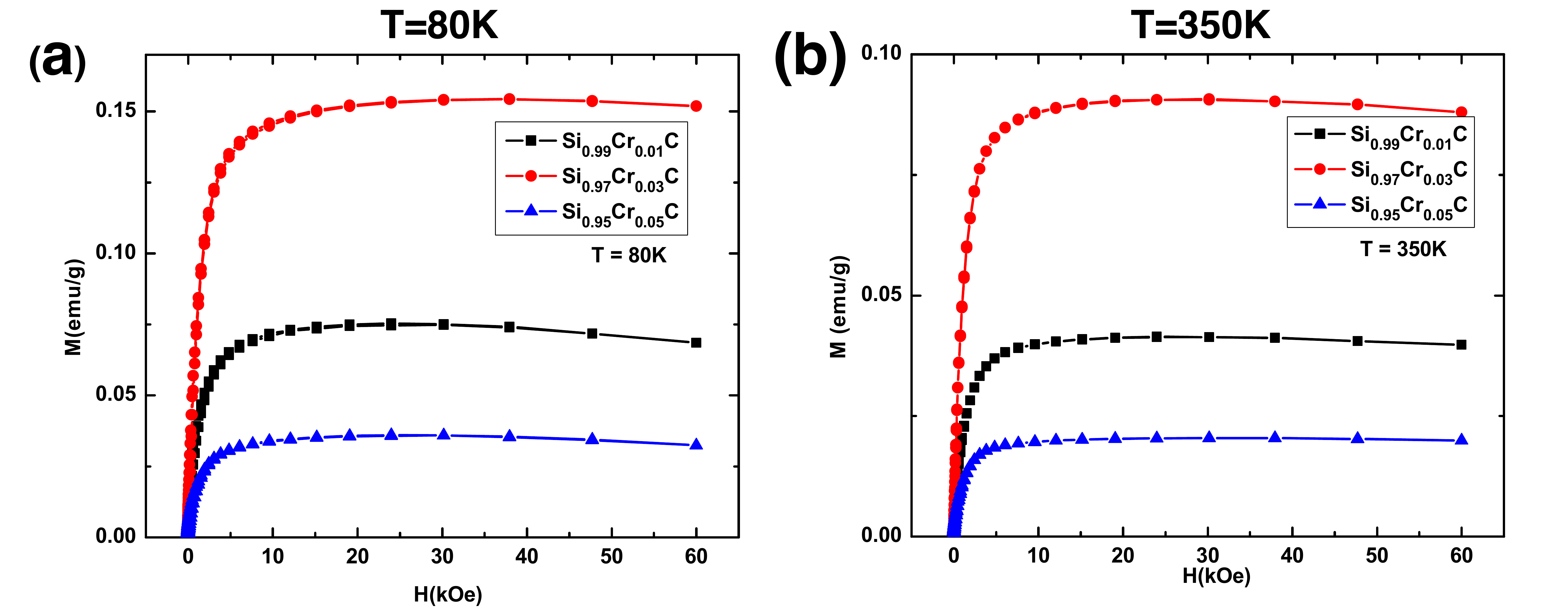}
\par\end{centering}
\caption{\label{fig:.BMP fit with Experimental data after subtraction para magnetic contribution at T =00003D  80K and 350K respectviely.}
BMP fit with experimental data after subtracting para magnetic contribution
showing ferromagnetic part of Cr doped 3C-SiC.}
\end{figure}

\twocolumngrid

\noindent without the presence of defects \citep{Pal2010,Podila2010}.
This is because, in case of short range interaction, it would align
in an anti-ferromagnetic manner. So the doped magnetic cation prevents
it from happening and helps to set a long range magnetic interaction
in the system. The blue shift of the Raman spectra clearly reveals
the formation of Si/C vacancies. The no of cation sites $V_{c}$ within
a sphere of radius $r_{H}$ ranges from 10 to 100 depending upon the
value of $r_{H}$. The lower value of measured magnetization in undoped
3C-SiC (10$^{-3}$emu/g) must have been due to defect induced origin
as has been mention in the literature. But with Cr$^{3+}$ ion doping,
magnetization value has increased in Cr doped 3C-SiC. Thus it seems
that both magnetic ions as well as defects are important elements
to attain high moment as well as high $T_{\text{C}}$ as observed
here in our sample. Fig \ref{fig:.BMP fit with Experimental data after subtraction para magnetic contribution at T =00003D  80K and 350K respectviely.}(a,
b) shows ferromagnetic part contributed from BMP percolation after
subtracting the paramagnetic part \citep{Wang2015b}. Magnetic moment
of 3\% Cr doped sample is high as compared to 5\% Cr doped sample.
A plausible reason can be an increase in Cr concentration, as it favors
the formation of more no of defects and hence more no of BMPs. But
further increase in Cr concentration (5\%) may favour the initiative
of exchange interaction and reduce the magnetization as the distance
between Cr-Cr ions decreases.
\begin{singlespace}

\section{Conclusion}
\end{singlespace}

In this study, for the first time, we have reported room temperature
ferromagnetism in Cr doped 3C-SiC and found a noble method to determine
the charge carrier density by using the LOPC model. The X-band EPR
spectra of Cr doped 3C-SiC reveals multi valence states of chromium.
The Q-band EPR spectra clearly reveals the presence of Cr$^{3+}$
ions in the system. The non monotonous variation of Raman shift and
phonon life time can be explained by invoking phonon - carrier scattering
process (LOPC model). Room Temperature FM has been observed in Cr
doped 3C-SiC. The observed FM could be explained on the basis of intrinsic
exchange interaction of Cr ions and V$_{\text{Si}}$, V$_{\text{c}}$
defects present in the system. Both Cr$^{3+}$ ions, as well as defects,
play a decisive role to attain a high temperature long range magnetic
order. Substitution of Cr$^{3+}$ in place of Si$^{4+}$ enhances
the defect concentration and thereby sets up long range magnetic order
in the system. But the excess doping of Cr ion in the host matrix
introduces antiferromagnetic order and leads to the suppression of
the ferromagnetic interaction caused by BMP percolation. Solubility
limit of Cr in 3C-SiC has been found to be 3\%.
\begin{acknowledgments}
\begin{singlespace}
\noindent One of the authors, Gyanti Prakash Moharana, thanks IIT
Madras for the financial support. Rahul Kothari sincerely acknowledges
the Institute Post Doctoral Fellowship of IIT Ma-
\end{singlespace}
\end{acknowledgments}

\newpage
\onecolumngrid

\begin{table}
\subfloat[BMP Fitting Parameters of Si$_{0.99}$Cr$_{0.01}$C]{\begin{centering}
\begin{tabular}{ccccccc}
\hline 
T(K) & M$_{\text{eff}}$$\times10^{-17}$ (emu) &  $\chi_{m}\times10^{-6}$(emu/gOe)  & N$\times10^{17}$(cm$^{-3}$) & M$_{s}$(emu /g) & H$_{C}$(Oe) & M$_{r}$(m emu/g)\tabularnewline
\hline 
350 & 7.65 & 1.30 & 0.016 & 0.04 & 64 & 0.002\tabularnewline
80 & 1.79 & 4.17 & 0.013 & 0.07 & 180 & 0.011\tabularnewline
5 & 104 & 0.11 & 236 & 0.77 & 300 & 0.023\tabularnewline
\hline 
\end{tabular}
\par\end{centering}
}

\subfloat[BMP Fitting Parameters of Si$_{0.97}$Cr$_{0.03}$C]{\begin{centering}
\begin{tabular}{ccccccc}
\hline 
T(K) & M$_{\text{eff}}$$\times10^{-17}$ (emu) &  $\chi_{m}\times10^{-6}$(emu/gOe)  & N$\times10^{17}$(cm$^{-3}$) & M$_{s}$(emu /g) & H$_{C}$(Oe) & M$_{r}$(m emu/g)\tabularnewline
\hline 
350 & 8.7 & 70.02 & 0.033 & 0.09 & 8 & 0.006\tabularnewline
80 & 1.92 & 3.09 & 0.025 & 0.15 & 71 & 0.009\tabularnewline
5 & 328 & 0.12 & 43.320 & 0.44 & 268 & 0.036\tabularnewline
\hline 
\end{tabular}
\par\end{centering}
}

\subfloat[BMP Fitting Parameters of Si$_{0.95}$Cr$_{0.05}$C]{\begin{centering}
\begin{tabular}{ccccccc}
\hline 
T(K) & M$_{\text{eff}}$$\times10^{-17}$ (emu) &  $\chi_{m}\times10^{-6}$(emu/gOe)  & N$\times10^{17}$(cm$^{-3}$) & M$_{s}$(emu /g) & H$_{C}$(Oe) & M$_{r}$(m emu/g)\tabularnewline
\hline 
350 & 8.15 & 536 & 0.008 & 0.02 & 70 & 0.001\tabularnewline
80 & 1.78 & 2.35 & 0.065 & 0.03 & 150 & 0.004\tabularnewline
5 & 780 & 5.85 & 390.210 & 0.94 & 210 & 0.013\tabularnewline
\hline 
\end{tabular}
\par\end{centering}
}

\caption{\label{tab:BMP fit of Cr doped 3C-SiC} BMPs Fitting parameters extracted
from experimental data by using BMP model.}
\end{table}

\twocolumngrid

\noindent dras. We greatly acknowledge DST-SAIF, IIT Madras for VSM
and EPR measurements. We thank Prof. Sankaran Subramanian (Adjunct
Professor \& INSA Senior Scientist, Department of Chemistry and Sophisticated
Analytical Instrument Facility) for useful discussion on the EPR data
analysis.

\bibliographystyle{unsrt}
\bibliography{bibliography}

\end{document}